\documentclass[aps,prb,preprint]{revtex4}
\usepackage{graphicx,color}
\usepackage{dcolumn}
\usepackage{bm}
\usepackage{amsmath,amsthm,amssymb}

\begin{document}

\title{
Calculation of current-induced torque from spin continuity equation
}

\author{Gen Tatara$^{1}$ and Peter Entel$^{2}$}
\affiliation{%
$^{1}$Department of Physics, Tokyo Metropolitan University,
Hachioji, Tokyo 192-0397, Japan
\\
$^{2}$Physics Department,
University of Duisburg-Essen,
47048 Duisburg, Germany
}

\date{\today}

\begin{abstract}
Current-induced torque is formulated based on the spin continuity equation.
The formulation does not rely on the assumption of separation of local spin and charge degrees of freedom, in contrast to approaches based on the $s$-$d$ model or mean-field approximation of itinerant ferromagnetism. 
This new method would be thus useful for the estimation of torques in actual materials 
by first-principles calculations.
As an example, the formalism is applied to 
the adiabatic limit of the $s$-$d$ model in order to obtain the analytical expression for torques and corresponding $\beta$ terms arising from spin relaxation due to spin-flip scattering and spin-orbit interaction. 

\end{abstract}


\maketitle
\def\average#1{\left\langle {#1} \right\rangle}
\def\ddelo#1{\frac{d^2}{d #1^2}}
\def\ddel#1#2{\frac{d^2 #1}{d #2 ^2}}
\def\ddelpo#1{\frac{\partial^2}{\partial #1^2}}
\def\delo#1{\frac{d}{d #1}}
\def\delpo#1{\frac{\partial}{\partial #1}}
\def\deldo#1{\frac{\delta}{\delta #1}}
\def\del#1#2{\frac{d #1}{d #2}}
\def\delp#1#2{\frac{\partial #1}{\partial #2}}
\def\deld#1#2{\frac{\delta #1}{\delta #2}}
\def\vvec#1{\stackrel{{\leftrightarrow}}{#1}} 
\def\vec2#1#2{\left(\begin{array}{c} #1 \\ #2 \end{array}\right)}
\def\vec3#1#2#3{\left(\begin{array}{c} #1 \\ #2 \\ #3 \end{array}\right)}
\def\listitem#1{\begin{itemize}\item #1 \end{itemize}}
\newcommand{\lt}{\left}
\newcommand{\rt}{\right}
\newcommand{\nablarl}{\stackrel{\leftrightarrow}{\nabla}}
\newcommand{\nablal}{\stackrel{\leftarrow}{\nabla}}
\newcommand{\nablar}{\stackrel{\rightarrow}{\nabla}}
\newcommand{\band}{n}
\newcommand{\ee}{{\rm ee}}
\newcommand{\Hee}{H_{\ee}}
\newcommand{\Hsr}{H_{\rm sr}}
\newcommand{\lamso}{\lambda_{\rm so}}
\newcommand{\nso}{n_{\rm so}}
\newcommand{\pvh}{\hat{\bm{p}}}
\newcommand{\torquen}{\torque_{\rm n}}
\newcommand{\kfu}{{k_{F+}}}
\newcommand{\kfd}{{k_{F-}}}
\newcommand{\tauso}{\tau^{\rm (so)}}
\newcommand{\Gtil}{\tilde{G}}
\newcommand{\Jsd}{J_{sd}}
\newcommand{\nel}{n_{\rm e}}
\newcommand{\Simptil}{\tilde{S}_{\rm imp}}
\def\Eqref#1{Eq. (\ref{#1})}
\def\Eqsref#1#2{Eqs. (\ref{#1})(\ref{#2})}
\newcommand{\adag}{{a^{\dagger}}}
\newcommand{\Area}{A}
\newcommand{\Av}{{\bm A}}
\newcommand{\Avem}{{\bm A}_{\rm em}}
\newcommand{\Ams}{{\rm A/m}^2}
\newcommand{\Aph}{A^{\phi}}
\newcommand{\Ath}{A^{\theta}}
\newcommand{\Aphv}{\Av^{\phi}}
\newcommand{\Athv}{\Av^{\theta}}
\newcommand{\Az}{{A^{z}}}
\newcommand{\Bv}{{\bm B}}
\newcommand{\Bz}{{B_z}}
\newcommand{\Bc}{B_{\rm c}}
\newcommand{\Bveff}{{\bm B}_{\rm eff}}
\newcommand{\Bve}{{\bm B}_{\rm e}}
\newcommand{\betasf}{{\beta_{\rm sf}}}
\newcommand{\betaso}{{\beta_{\rm so}}}
\newcommand{\betasr}{{\beta_{\rm sr}}}
\newcommand{\cdag}{{c^{\dagger}}}
\newcommand{\chiz}{\chi^{(0)}}
\newcommand{\chio}{\chi^{(1)}}
\newcommand{\chitilo}{\tilde{\chi}^{(1)}}
\newcommand{\chitilz}{\tilde{\chi}^{(0)}}
\newcommand{\ckv}{c_{\kv}}
\newcommand{\ckvs}{c_{\kv\sigma}}
\newcommand{\dels}{{s_0}}
\newcommand{\dx}{{d^3 x}}
\newcommand{\DOS}{{\nu}}
\newcommand{\DOSV}{{N(0)}}
\newcommand{\DOSom}{{\nu_\omega}}
\newcommand{\ef}{{\epsilon_F}}
\newcommand{\eF}{{\epsilon_F}}
\newcommand{\ekv}{\epsilon_{\kv}}
\newcommand{\ekvs}{\epsilon_{\kv\sigma}}
\newcommand{\Ev}{{\bm E}}
\newcommand{\ev}{{\bm e}}
\newcommand{\evth}{{\bm e}_{\theta}}
\newcommand{\evph}{{\bm e}_{\phi}}
\newcommand{\evs}{{\bm n}}
\newcommand{\evsz}{{\evs}_0}
\newcommand{\evsph}{(\evph\times\evz)}
\newcommand{\evz}{{\bm e}_{z}}
\newcommand{\fl}{{\eta}}
\newcommand{\fltil}{\tilde{\eta}}
\newcommand{\flitil}{\tilde{\fl_{\rm I}}}
\newcommand{\flrtil}{\tilde{\fl_{\rm R}}}
\newcommand{\fli}{{\fl_{\rm I}}}
\newcommand{\flr}{{\fl_{\rm R}}}
\newcommand{\fkvs}{f_{\kv\sigma}}
\newcommand{\fo}{{f(\omega)}}
\newcommand{\fpo}{{f'(\omega)}}
\newcommand{\fpin}{{f_{\rm pin}}}
\newcommand{\fe}{{f_{\rm e}}}
\newcommand{\fna}{{f_{\rm ref}}}
\newcommand{\Fe}{F}
\newcommand{\Fev}{\Fv}
\newcommand{\Fbeta}{{F^{\beta}}}
\newcommand{\Fena}{{F^{\rm rf}}}
\newcommand{\Fvna}{{\Fv^{\rm ref}}}
\newcommand{\Fna}{{F^{\rm ref}}}
\newcommand{\Fad}{{F^{\rm ad}}}
\newcommand{\Foad}{{\delta F^{\rm (1)ad}}}
\newcommand{\Fzad}{{F^{\rm (0)ad}}}
\newcommand{\Fv}{{\bm F}}
\newcommand{\Fo}{{F^{(1)}}}
\newcommand{\Fw}{F_{\rm w}}
\newcommand{\Fz}{{F^{(0)}}}
\newcommand{\Fdel}{{\delta \Fo}}
\newcommand{\gv}{{\bm g}}
\newcommand{\gr}{g^{\rm r}}
\newcommand{\ga}{g^{\rm a}}
\newcommand{\gless}{g^{<}}
\newcommand{\Gr}{G^{\rm r}}
\newcommand{\Ga}{G^{\rm a}}
\newcommand{\Gless}{G^{<}}
\newcommand{\ggtr}{g^{>}}
\newcommand{\gap}{\Delta_{\rm sw}}
\newcommand{\gammap}{\gamma_{+}}
\newcommand{\gammam}{\gamma_{-}}
\newcommand{\hf}{\frac{1}{2}}
\newcommand{\HA}{{H_{A}}}
\newcommand{\HB}{H_{B}}
\newcommand{\He}{H_{\rm e}}
\newcommand{\Heff}{H_{\rm eff}}
\newcommand{\Hem}{H_{\rm em}}
\newcommand{\Hex}{H_{\rm ex}}
\newcommand{\Himp}{H_{\rm imp}}
\newcommand{\Hint}{H_{\rm int}}
\newcommand{\HR}{{H_{\rm R}}}
\newcommand{\Hs}{{H_{\rm S}}}
\newcommand{\Hsf}{{H_{\rm sf}}}
\newcommand{\Hso}{{H_{\rm so}}}
\newcommand{\Hst}{{H_{\rm ST}}}
\newcommand{\hbarinv}{\frac{1}{\hbar}}
\renewcommand{\Im}{{\rm Im}}
\newcommand{\intinf}{\int_{-\infty}^{\infty}}
\newcommand{\intom}{\int \frac{d\omega}{2\pi}}
\newcommand{\intx}{\int {d^3x}}
\newcommand{\iv}{\bm{i}}
\newcommand{\Iv}{\bm{I}}
\newcommand{\Js}{{J_{\rm s}}}
\newcommand{\js}{j_{\rm s}}
\newcommand{\jsc}{{j_{\rm s}^{\rm c}}}
\newcommand{\jsv}{\bm{j}_{\rm s}}
\newcommand{\jc}{j_{\rm c}}
\newcommand{\jci}{{{j}_{\rm c}^{\rm i}}}
\newcommand{\jce}{{{j}_{\rm c}^{\rm e}}}
\newcommand{\jatil}{{\tilde{j}_{\rm a}}}
\newcommand{\jctil}{{\tilde{j}_{\rm c}}}
\newcommand{\jcitil}{{\tilde{j}_{\rm c}^{\rm i}}}
\newcommand{\jcetil}{{\tilde{j}_{\rm c}^{\rm e}}}
\newcommand{\jstil}{{\tilde{j}_{\rm s}}}
\newcommand{\jtil}{{\tilde{j}}}
\newcommand{\jv}{\bm{j}}
\newcommand{\kB}{{k_B}}
\newcommand{\kb}{{k_B}}
\newcommand{\kv}{{\bm k}}
\newcommand{\kvp}{{\kv}'}
\newcommand{\kpq}{{k+\frac{q}{2}}}
\newcommand{\kmq}{{k-\frac{q}{2}}}
\newcommand{\kvpq}{{\kv}+\frac{\qv}{2}}
\newcommand{\kvmq}{{\kv}-\frac{\qv}{2}}
\newcommand{\kvppq}{{\kvp}+\frac{\qv}{2}}
\newcommand{\kvpmq}{{\kvp}-\frac{\qv}{2}}
\newcommand{\kf}{{k_F}}
\newcommand{\kF}{{k_F}}
\newcommand{\kfpm}{{k_{F\pm}}}
\newcommand{\kfmp}{{k_{F\mp}}}
\newcommand{\kom}{{k_\omega}}
\newcommand{\Kp}{{K_\perp}}
\newcommand{\ktil}{\tilde{k}}
\newcommand{\lams}{{\lambda_{\rm s}}}
\newcommand{\lamv}{{\lambda_{\rm v}}}
\newcommand{\lamz}{{\lambda_{0}}}
\newcommand{\Le}{{L_{\rm e}}}
\newcommand{\Lez}{{L_{\rm e}^0}}
\newcommand{\Lb}{{L_{\rm B}}}
\newcommand{\Ldw}{{L_{\rm dw}}}
\newcommand{\Ls}{{L_{\rm S}}}
\newcommand{\Lsw}{{L_{\rm sw}}}
\newcommand{\Lswdw}{{L_{\rm sw-dw}}}
\newcommand{\Linv}{{\frac{1}{L}}}
\newcommand{\mv}{{\bm m}}
\newcommand{\Mv}{{\bm M}}
\newcommand{\Mw}{{M_{\rm w}}}
\newcommand{\mus}{{g\mu_{B}}}
\newcommand{\mub}{\mu_B}
\newcommand{\muB}{\mu_B}
\newcommand{\Ne}{N_{\rm e}}
\newcommand{\nv}{{\bm n}}
\newcommand{\nimp}{n_{\rm imp}}
\newcommand{\nvortex}{n_{\rm v}}
\newcommand{\nz}{{n}_0}
\newcommand{\om}{{\omega}}
\newcommand{\Omegatil}{{\tilde{\Omega}}}
\newcommand{\Omegap}{{{\Omega'}}}
\newcommand{\Omegaptil}{{\tilde{\Omegap}}}
\newcommand{\Omz}{{{\Omega_0}}}
\newcommand{\ompOmz}{{\omega+\frac{\Omz}{2}}}
\newcommand{\ommOmz}{{\omega-\frac{\Omz}{2}}}
\newcommand{\phiz}{{\phi_0}}
\newcommand{\PhiB}{{\Phi_{\rm B}}}
\newcommand{\Ptil}{{\tilde{P}}}
\newcommand{\pv}{{\bm p}}
\newcommand{\qv}{{\bm q}}
\newcommand{\qtil}{{\tilde{q}}}
\newcommand{\ra}{\rightarrow}
\renewcommand{\Re}{{\rm Re}}
\newcommand{\rhow}{{\rho_{\rm w}}}
\newcommand{\rhos}{{\rho_{\rm s}}}
\newcommand{\rhoS}{{\rho_{\rm s}}}
\newcommand{\rhoxy}{{\rho_{xy}}}
\newcommand{\RS}{{R_{\rm S}}}
\newcommand{\Rw}{{R_{\rm w}}}
\newcommand{\rv}{{\bm r}}
\newcommand{\Rv}{{\bm R}}
\newcommand{\sigmav}{{\bm \sigma}}
\newcommand{\se}{{s}}
\newcommand{\sev}{{\bm \se}}
\newcommand{\sgn}{{\rm sgn}}
\newcommand{\sz}{{s}_0}
\newcommand{\sv}{{{\bm s}}}
\newcommand{\seth}{{\se}_\theta}
\newcommand{\seph}{{\se}_\phi}
\newcommand{\sez}{{\se}_z}
\newcommand{\so}{\lambda}
\newcommand{\spol}{{M}}
\newcommand{\Jex}{{J_{\rm ex}}}
\newcommand{\svtil}{\tilde{\bm s}}
\newcommand{\stil}{\tilde{s}}
\newcommand{\stilz}{\stil_{z}}
\newcommand{\stilpm}{\stil^{\pm}}
\newcommand{\stilpmz}{\stil^{\pm(0)}}
\newcommand{\stilpma}{\stil^{\pm(1{\rm a})}}
\newcommand{\stilpmb}{\stil^{\pm(1{\rm b})}}
\newcommand{\stilpmo}{\stil^{\pm(1)}}
\newcommand{\stilpara}{\stil_{\parallel}}
\newcommand{\stilperp}{\stil_{\perp}}
\newcommand{\Simpv}{{{\bm S}_{\rm imp}}}
\newcommand{\Simp}{{S_{\rm imp}}}
\newcommand{\Stot}{{S_{\rm tot}}}
\newcommand{\Sh}{{\hat {S}}}
\newcommand{\Svh}{{\hat {\Sv}}}
\newcommand{\Sv}{{{\bm S}}}
\newcommand{\Svz}{{{\bm S}_0}}
\newcommand{\sumx}{{\int \frac{d^3x}{a^3}}}
\newcommand{\sumk}{{\sum_{k}}}
\newcommand{\sumkv}{{\sum_{\kv}}}
\newcommand{\sumom}{\int\frac{d\omega}{2\pi}}
\newcommand{\sumOm}{\int\frac{d\Omega}{2\pi}}
\newcommand{\sumomOm}{\int\frac{d\omega}{2\pi}\int\frac{d\Omega}{2\pi}}
\newcommand{\sumqv}{{\sum_{\qv}}}
\newcommand{\thickness}{{d}}
\newcommand{\thetaz}{{\theta_0}}
\newcommand{\tr}{{\rm tr}}
\newcommand{\Tc}{{T_{C}}}
\newcommand{\Torqv}{{\bm \tau}}
\newcommand{\torque}{{\tau}}
\newcommand{\torquev}{{\bm \torque}}
\newcommand{\Torquev}{{\bm \tau}}
\newcommand{\torqueve}{{\torquev_{\rm e}}}
\newcommand{\torquee}{{\torque_{\rm e}}}
\newcommand{\torquew}{{\torque_{\rm w}}}
\newcommand{\tautil}{{\tilde{\tau}}}
\newcommand{\taup}{\tau_{+}}
\newcommand{\taum}{\tau_{-}}
\newcommand{\tauw}{\tau_{\rm w}}
\newcommand{\tausf}{\tau_{\rm sf}}
\newcommand{\thetast}{\theta_{\rm st}}
\newcommand{\ttil}{{\tilde{t}}}
\newcommand{\tz}{{t_0}}
\newcommand{\vc}{{v^{\rm c}}}
\newcommand{\ve}{{v_{\rm e}}}
\newcommand{\vv}{\bm{v}}
\newcommand{\vs}{{v_{\rm s}}}
\newcommand{\vsv}{{\vv_{\rm s}}}
\newcommand{\vf}{{v_F}}
\newcommand{\vimp}{v_{\rm imp}}
\newcommand{\vi}{{v_{\rm i}}}
\newcommand{\Vso}{V_{\rm so}}
\newcommand{\vtil}{{{v_0}}}
\newcommand{\Vpin}{{V}_{\rm pin}}
\newcommand{\Vinv}{\frac{1}{V}}
\newcommand{\vz}{{v_0}}
\newcommand{\Vz}{{V_0}}
\newcommand{\Vztil}{{\tilde{V_0}}}
\newcommand{\Ws}{{W_{\rm S}}}
\newcommand{\Xtil}{{\tilde{X}}}
\newcommand{\xv}{{\bm x}}
\newcommand{\Xv}{{\bm X}}
\newcommand{\xvp}{{\bm x}_{\perp}}
\newcommand{\xw}{{z}}
\newcommand{\Xz}{{X_0}}
\newcommand{\Zs}{Z_{\rm S}}
\newcommand{\ztil}{u}


\section{Introduction}

Spin transfer torque is a torque acting on local spins
as a result of an applied current. 
Such a torque has been discussed mostly based on $s$-$d$ type of exchange interaction \cite{TK04,Zhang04,Waintal04,KTS06,Kohno07,Tatara_cd08}
after the pioneering works by Berger\cite{Berger78,Berger84} and Slonczewski\cite{Slonczewski96}.
In $s$-$d$ models, the conduction electrons and localized spins are discriminated, and therefore the transfer of spin angular momentum between those two degrees of freedom occurs.
However, in reality, this separation of degrees of freedom is not always so obvious, since in an itinerant picture all electronic bands contribute to both conduction and magnetism with different weights.
Thus, the formulation of spin torques based on the $s$-$d$ picture is an approximation, and this is a serious problem when one tries to evaluate current-induced torques in actual materials.
For trustful estimates, formulations beyond the simple $s$-$d$ separation is certainly required.
Such a formalism can be combined with first-principles calculations without any artificial assumption and would be useful for realistic estimates of current-induced torques and efficiency of current-induced switching.
The aim of this paper is to develop a new calculational scheme satisfying these requirements 
based on the spin continuity equation. 

Theoretical determination of current-induced torques is difficult  even in the simplest case of $s$-$d$ model when spin relaxation and non-adiabaticity is present\cite{Tserkovnyak06,KTS06,Kohno07,TKSLL07,Duine07,Thorwart07,Piechon07,TKS08}.
So far, very few studies on the effect of spin relaxation due to spin flip scattering by magnetic impurities have been done microscopically\cite{KTS06,Kohno07,Duine07}.
In the $s$-$d$ formalism, the current-induced torque is represented as the effective field due to the spin polarization of the conduction electron, $\sev$.
The torque is therefore given as
$\torquev^{(sd)}=-\Jsd \Sv\times\sev$, where
$\Sv$ is the localized ($d$) electron spin and $\Jsd$ is the exchange interaction constant.
Microscopic calculation using linear response theory\cite{KTS06,Kohno07} revealed, in agreement with phenomenolocigal result\cite{Zhang04}, that spin-flip interaction of conduction electrons with random impurities induces a torque perpendicular to the spin-transfer torque (called $\beta$ terms\cite{Thiaville05}). 
The torque is written as
\begin{equation}
\torquev^{(\beta)}=-\beta \frac{P}{eS^2}(\Sv\times(\jv\cdot\nabla)\Sv),
\end{equation}
where $P$ is the spin polarization of the current and $\jv$ is the current density.
The coefficient $\beta$ was calculated by 
summing over not a few Feynman diagrams representing self-energy and vertex corrections\cite{KTS06,Kohno07,TKS08}.

The case of itinerant ferromagnetism was studied by Tserkovnyak \textit{et al.}\cite{Tserkovnyak06} and Duine \textit{et al.}\cite{Duine07}.
They introduced the magnetization as a mean-field expectation value of itinerant electron spin, and thus the models considered were effectivelly the $s$-$d$ model. 
Tserkovnyak \textit{et al.} considered a kinetic equation for the spin density with a consistency condition for the magnetization, but the spin dephasing term was introduced phenomenologically.
Duine \textit{et al.} estimated the torques by calculating the effective action for the magnetization fluctuation, which has been assumed to be of small amplitude.
Within the mean-field treatment, the toruqe in the itinerant case turned out to be exactly the same as that of the $s$-$d$ model\cite{Duine07,KTS06}.

It has recently been noticed that the coefficient $\beta$ is  very important for the realization of highly efficient magnetization switching by the current\cite{Zhang04,Thiaville05,TTKSNF06,TKS08}.
First, it affects the threshold current, and the intrinsic pinning threshold is replaced by an extrinsic one, which is usually lower than the intrinsic one.
Second, it results in a terminal speed of the wall,
$v\propto \frac{\beta}{\alpha}j$, which can exceed the pure spin-transfer speed limit if $\frac{\beta}{\alpha}$ is large ($\alpha$ is Gilbert damping parameter).
Third, the deformation of the wall depends on $\beta$.
When $\beta\sim \alpha$, deformation is suppressed and weak dissipation may be expected\cite{Heyne08}.
Experimental studies of the value of $\beta$ have recently been carried out. 
Significant wall deformation observed in permalloy indicated that $\beta\neq \alpha$\cite{Heyne08}.
Thomas \textit{et al.}\cite{Thomas06} found for permalloy that the observed wall speed corresponds to $\beta\sim 8\alpha$.
Therefore, determination of $\beta$ is of particular importance for device applications.

In this paper, we will present a microscopic calculation scheme  different from the $s$-$d$ formalism\cite{Berger78,Berger84}.
The idea is simply to use the continuity equation of spin, and thus the formulation is not necessarily based on the $s$-$d$ interaction picture. 
The formalism turns out to be quite powerful in particular, for the determination of spin relaxation effect, $\beta$.
The continuity equation which we consider is essentially the kinetic equation discussed by Tserkovnyak \textit{et al.}\cite{Tserkovnyak06}, but all observables have been microscopically defined and can be calculated using our formalism.
For instance, spin dephasing time introduced phenomenolocigally in Ref. \cite{Tserkovnyak06} is represented by the spin source term (${\cal T}$) defined by Green's function in our formalism.
Microscopic details of this term ${\cal T}$ turn out to be essential in determining the spin-relaxation induced torque.

Our scheme is applicable also to the $s$-$d$ model or mean-field approximation of itinerant ferromagnetism.
We will use our formalism to  obtain the analytical expression of the torques arising from both spin-flip scattering and spin-orbit interaction arising from the impurities in the $s$-$d$ model in the adiabatic limit. 
In the present formalism, the number of contributing diagrams  is  less than the number of diagrams used in the $s$-$d$ exchange formalism\cite{KTS06,Kohno07}, and thus the calculation is easier.

\section{Formalism}

The spin density $\sev$ of the total system is defined as
the expectation value of conduction electron spin, summed over all bands $\band$ as
\begin{align}
\se^\alpha(\xv,t) &\equiv 
\sum_{\band} \average{c^\dagger_{\band}(\xv,t)\sigma^\alpha c_{\band}(\xv,t)}. 
\end{align}
It satisfies the equation, 
$\hbar\dot{\se}^\alpha 
=i(
\average{[H,c^\dagger_{\band}]\sigma^\alpha c_{\band}}
+\average{c^\dagger_{\band}\sigma^\alpha [H,c_{\band}]}
)$, where $H$ is total Hamiltonian. 
We assume that $H$ consists of free part, 
spin relaxation part, $\Hsr$, as
$H=\intx \sum_{\band}\frac{\hbar^2}{2m}|\nabla c_{\band}|^2
+\Hsr$. 
Then the continuity equation is obtained as 
\begin{align}
\hbar\dot{\se}^\alpha &= - \frac{1}{e} \nabla\cdot \jsv^\alpha 
 +{\cal T}^\alpha,
   \label{cont}
\end{align}
where $e$ represents the electron charge.
Here the spin current $\jsv$ is defined by the free part as
\begin{align}
{\js}_\mu^\alpha &\equiv 
-\frac{ie\hbar}{2m}
\sum_{\band}\average{c^\dagger_{\band}(\xv,t)\stackrel{\leftrightarrow}{\nabla_\mu} \sigma^\alpha c_{\band}(\xv',t)},
\end{align}
and the spin source (or sink) 
$\bf{\cal T}$ is a contribution arising from spin relaxation and interaction, i.e., 
\begin{align}
{\bf{\cal T}}^\alpha
\equiv i \sum_{\band}
(\average{[\Hsr,c^\dagger_{\band}]\sigma^\alpha c_{\band}}
+\average{c^\dagger_{\band}\sigma^\alpha [\Hsr,c_{\band}]}
).
\end{align}
The continuity equation (\ref{cont}) is sufficient to calculate the torque acting on the spin.
Actually, the equation is equivalent to the equation of motion of spin,
$\hbar\dot{\sev}=\torquev$, where $\torquev$ represents the total torque acting on the spin.
The torque is thus simply given by
\begin{align}
\torque^\alpha &= - \frac{1}{e} \nabla\cdot \jsv^\alpha 
  +{\cal T}^\alpha 
 \label{torquedef}.
\end{align}
 
Note that the continuity equation describes the time-dependence of the spin density, and therefore the right-hand side of Eqs. (\ref{cont}) and (\ref{torquedef}) is uniquely defined even in the presence of spin relaxation, where the spin current can be defined in several different ways, see Ref. \cite{Murakami04}.
In the context of spin Hall effect, the continuity equation (\ref{cont}) was used to obtain \textit{proper} definition of spin current and to explore transport properties  \cite{Culcer04,Shi06,Zhang08}.
Concerning the current-induced torques, 
the equation (\ref{torquedef}) has been so far applied only in the absence of spin relaxation term, where the torque is given by the divergence of the spin current 
\cite{Caroli71,Wang08}.
The main aim of this paper is to study the spin relaxation contribution, 
${\cal T}$.

Let us look explicitly at the continuity equation in case of spin relaxation due to spin impurities and spin-orbit interaction, $\Hsr=\Hsf+\Hso$. 
Spin flip interaction is described by
\begin{align}
\Hsf &= \vs \intx \sum_{\band} 
 \Simpv(\xv) \cdot (c^\dagger_{\band} \sigmav c_{\band}), 
\end{align}
where $\vs$ is a constant,
$\Simpv(\xv)\equiv
\sum_{i}^{\nimp}{\Simpv}_i\delta(\xv-\Rv_i)$,  
 ${\Simpv}_i$ represents the impurity spin at $\xv=\Rv_i$ and $\nimp$ denotes the number of impurity spins.
The spin-orbit interaction is written as
\begin{align}
\Hso &=  -\frac{i}{2}\lamso \intx  \sum_{ijkl}
\epsilon_{ijk} \nabla_j \Vso^{(l)}(\xv)
 (c^\dagger_{\band} \sigma^l \nablarl_k c_{\band}),
\end{align}
where the potential $\Vso^{(l)}$  is here assumed to arise from random impurities and depends on the spin direction ($l$).
 
The spin-relaxation torque is given by a sum of contributions from spin-flip and spin-orbit interaction as 
${\cal T}^\alpha ={\cal T}^\alpha_{\rm sf} +{\cal T}^\alpha_{\rm so} $, where 
\begin{align}
{\cal T}^\alpha_{\rm sf} (\xv)  & = 
 2\vs \sum_{\beta\gamma} \epsilon_{\alpha\beta\gamma}
 \average{{S^{\beta}_{\rm imp}} \se^\gamma}_{\rm i} 
\label{Tsr}  \\
{\cal T}^\alpha_{\rm so} (\xv) & = 
 - 2m\lamso \sum_{\beta\gamma\mu\nu} \epsilon_{\alpha\beta\gamma}
 \epsilon_{\mu\nu\beta} 
\average{\nabla_\mu \Vso^{(\gamma)}(\xv) {\js}^\gamma _{\rm \nu}}_{\rm i} 
\label{Tso} . 
\end{align}
The average over random impurity spins and spin-orbit potential is represented by $\average{\ }_{\rm i}$.

All the terms on the right-hand side of torque Eqs. (\ref{torquedef}), (\ref{Tsr}) and (\ref{Tso}) are written in terms of local spin density and local spin current, and so the torque acting on the spin is calculated by estimating the spin density and the spin current.
This representation of the spin torque applies to any spin relaxation processes and interaction, and is directly calculable without assuming separation of spin and charge degrees of freedoms.
Equations (\ref{torquedef}), (\ref{Tsr}) and (\ref{Tso}) are thus a suitable starting point for realistic estimates based on first-principles calculations.
This is the essential point of this paper (although, 
\textit{ab initio} calculations using the present formalism still need 
to be undertaken).

\section{Application to the $s$-$d$ model}

In the latter part of the paper, we will apply this formulation to estimate the current-induced torques in the adiabatic limit (i.e., slowly varying magnetization compared with conduction electron motion) to show the validity and usefulness of our formalism.
We will calculate the torque arising from the spin relaxation due to both the spin flip scattering and the spin-orbit interaction.
It is found that the torque is represented by the so called the $\beta$ term in both cases and values of corresponding $\beta$ are calculated. 
Out formulation is thus useful for both the analytical and the numerical studies.

We will now consider $s$-$d$ model with only one conduction band. 
(Please note that the assumption of separation of $s$ and $d$ electron here is simply for analytical demonstration and is not a requirement of present formulation. )
The $s$-$d$ interaction between a localized spin $\Sv$ and the conduction electrons is given by
\begin{equation}
\Hex \equiv -{\Jsd} \intx
  \Sv \cdot (c^\dagger \sigmav c).
\end{equation}
We describe the adiabatic limit by the standard local gauge transformation in the spin space, choosing the electron spin quantization axis along $\Sv(\xv,t)$ at each point.
A new electron operator $a\equiv ({a_+},{a_-})^{\rm t}$ 
(t denotes transpose) is defined as
$c(\xv,t)\equiv U(\xv,t) a(\xv,t),$ 
where $U$ is a $2\times2$ matrix which we further define as
$U(\xv,t)\equiv \mv\cdot\sigmav$,
$\mv$ being a real three-component unit vector
$\mv=\left(
\sin\frac{\theta}{2}\cos\phi,\sin\frac{\theta}{2}\sin\phi,\cos\frac{\theta}{2} \right)$.
The gauge field is written as
$A_\mu^\alpha \equiv (\mv\times\partial_\mu \mv)^\alpha$. 
Then the Hamiltonian of $a$-electrons is given by the free part, 
$\sum_{\kv\sigma} \epsilon_{\kv\sigma} a^\dagger_{\kv\sigma} a_{\kv\sigma}$
($\epsilon_{\kv\sigma} \equiv \ekv -\sigma\spol$, $\sigma=\pm$ represents the spin),  
$\HA$, describing the interaction with the SU(2) gauge field,  and $\Hem$, the interaction with the external electric field which drives the current\cite{TK04,TKS08}.
Here, we consider static local spins in the adiabatic limit, where the momentum transfered by the gauge field to conduction electrons is negligibly small (compared to $\kf$), and take into account the gauge field only in linear order. 
Then, the gauge interaction is given by\cite{TKS08}
\begin{align}
\HA
&=
\frac{\hbar^2}{m} \sum_{\qv} \sum_{\mu} k_\mu A_\mu^\alpha(-\qv) 
  a^\dagger_{\kv}\sigma_\alpha a_{\kv}  .
\end{align}
The applied electric field is represented by the interaction
\begin{eqnarray}
\Hem &=& \sum_\mu \frac{ie\hbar E_\mu}{m\Omz}e^{i\Omz t}
\sumkv \lt[ k_\mu  a^{\dagger}_{\kv} a_{\kv}
+\sum_{\alpha\qv} A^\alpha _{\mu}(\qv) a^{\dagger}_{\kv}\sigma^{\alpha} a_{\kv}
\rt] +O(E^2), \label{hemmod}
\end{eqnarray}
where $\Omz$ is the frequency of the field chosen as 
$\Omz\ra0$ at the end of calculation.

The spin current part of torque is calculated in the adiabatic limit as
\begin{align}
-\nabla\cdot \jsv^\alpha & \simeq  
-(\nabla_\mu \nv) {\js}_{\mu}. \label{stt}
\end{align}
Here, $\nv\equiv \Sv/S$ and, therefore, this contribution corresponds to the standard spin transfer torque.

\subsection{Torque from spin-flip scattering}

\begin{figure}[htb]
\begin{center}
\includegraphics[width=0.8\linewidth]{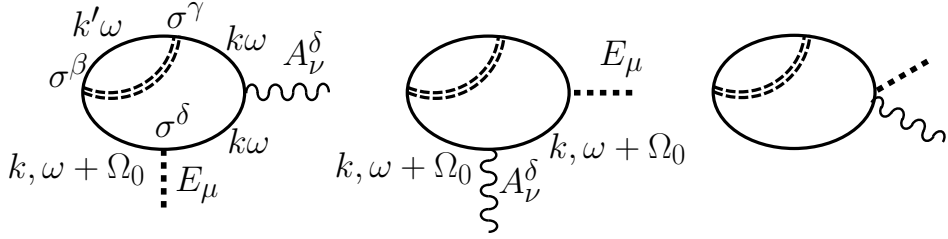}
\caption{The diagrammatic representation of $D_{\mu\nu}^{\beta\gamma\delta}$. 
Double dashed, dotted and wavy lines denote interaction with impurity spin, applied electric field $\Ev$ and gauge field $\Av$, respectively. 
\label{FIGdiag_sf}
}
\end{center}
\end{figure}

Let us turn to the spin relaxation part of the torque arising from spin impurities, i.e., Eq. (\ref{Tsr}).
(The effect of spin relaxation on the spin-current part can be shown to be simply due to modification of lifetime, $\tau$.)
Here, we assume that the impurity spins are influenced by a strong $s$-$d$ exchange field and write 
${\Simp}^{\alpha}(\xv)= R_{\alpha\beta}(\xv)\Simptil^{\beta}(\xv)$, where $\Simptil^{\alpha}$ represents impurity spin in the rotated frame, 
and 
\begin{align}
R_{\alpha\beta}\equiv 2m_\alpha m_\beta-\delta_{\alpha\beta},
\end{align}
is a rotation matrix.
Then the averaging is given by
$\average{ 
\tilde{S}_{{\rm imp}}^{\alpha}(\xv)
\tilde{S}_{{\rm imp}}^{\beta}(\xv') }_{\rm i}=
\frac{1}{3} \delta_{\alpha\beta}\delta(\xv-\xv')
\nimp{\overline{\Simp^2}}$, 
where $\nimp$ is the impurity spin concentration.
(Averaging taken with respect to $\Simp$ turns out to lead to essentially the same result as in the case of 
$\tilde{S}_{{\rm imp}}$.)
The spin source term is written as 
\begin{align}
{\cal T}^\alpha_{\rm sf}(\xv) & = 
- 2i\vs \sum_{\beta \gamma} 
F_{\alpha\beta\gamma}(\xv)
 \average{{\Simptil^{\beta}}(\xv) 
 \tr[\sigma^\gamma {\Gtil}^<_{\xv,\xv}]}_{\rm i},
\end{align}
where 
\begin{align}
F_{\alpha\beta\gamma}\equiv 
\sum_{\mu\nu}
\epsilon_{\alpha\mu\nu} R_{\mu\beta}R_{\nu\gamma},
\end{align}
and 
${\Gtil}^<_{\xv,\xv'}\equiv i\average{a^\dagger(\xv')a(\xv)}$
is the lesser component of the Green's function defined on Keldysh contour in the complex time.
To the lowest (second) order in $\vs$, we obtain after averaging over spin impurities
\begin{align}
{\cal T}^\alpha_{\rm sf}(\xv) 
&=  -i\frac{2}{3}\nimp\vs^2 \overline{\Simp^2} 
 \sum_{\beta\gamma\delta}\sum_{\mu\nu}
 F_{\alpha\beta\gamma}(\xv)
\tr[\sigma^\beta \Gtil^{(0)}_{\xv,\xv}\sigma^\gamma \Gtil^{(0)}_{\xv,\xv}]^<+O(\vs^4),\label{Tsf1}
\end{align}
where $\Gtil^{(0)}$ denotes Green's functions without impurity spins but including the gauge field $\Av$ and external electric field $\Ev$. 
Including these fields in linear order, we obtain 
\begin{align}
{\cal T}^\alpha_{\rm sf}(\xv)  
&=  - \frac{2e}{3m}\nimp \vs^2 \overline{\Simp^2}  \sum_{\beta\gamma\mu\nu}  F_{\alpha\beta\gamma}(\xv) E_{\mu} A^\delta_{\nu}(\xv)
D_{\mu\nu}^{\beta\gamma\delta},
\end{align}
where
\begin{align}
D_{\mu\nu}^{\beta\gamma\delta} &\equiv
\lim_{\Omz\ra0}\frac{1}{\Omz} \sumom \sum_{\kv\kv'} 
\tr\lt[ \sigma^\beta 
 \lt( 
\delta_{\mu\nu} 
g_{\kv'\omega} \sigma^\gamma g_{\kv\omega} \sigma^\delta g_{\kv\omega+\Omz}
\rt.\rt. \nonumber\\
&
\lt.\lt. 
+\frac{k_\mu k_\nu}{m}
  \lt\{ 
g_{\kv'\omega}\sigma^\gamma g_{\kv\omega}g_{\kv\omega+\Omz} \sigma^\delta g_{\kv\omega+\Omz} 
 +g_{\kv'\omega}\sigma^\gamma g_{\kv\omega}\sigma^\delta g_{\kv\omega} g_{\kv\omega+\Omz}  \rt\} 
\rt) \rt]^< +{\rm c.c.}. \label{Ddiag}
\end{align}
Here, the Green's function  $g_{\kv\omega}$ is the Fourier representation of free Green's function 
and $[\ ]^<$ denotes the lesser component.
They are diagonal in spin space, being defined in gauge-transformed space.
(Complex conjugates are denoted by c.c.)
Figure \ref{FIGdiag_sf} shows the contributions to $D_{\mu\nu}^{\beta\gamma\delta}$ diagrammatically.
The lesser component is calculated in standard manner in the limit of $\Omz\ra0$.
The first two diagrams of Fig. \ref{FIGdiag_sf} are simplified by use of partial integration over $\kv$ 
using 
$\frac{k_\mu}{m}(\ga_{\kv})^2
 =\frac{\partial}{\partial_{k_\mu}}\ga_{\kv}$ etc. 
These contributions are obtained as 
\begin{align}
{D_{\mu\nu}^{\beta\gamma\delta} }^{(1-2)}
& =
\lim_{\Omz\ra0} \sumom \sum_{\kv\kv'} 
\tr\lt[
f'(\omega) \frac{k_\mu k_\nu}{m}
\lt\{ \sigma^\beta \gr_{\kv'\omega}\sigma^\gamma    +\sigma^\gamma \ga_{\kv'\omega}\sigma^\beta \rt\}
(|\gr_{\kv\omega}|^2 \sigma^\delta \ga_{\kv\omega}
+ \gr_{\kv\omega} \sigma^\delta |\ga_{\kv\omega}|^2)  
    \rt.\nonumber\\
&
 +\delta_{\mu\nu}\lt[ 
\frac{f(\omega)}{2} 
\lt\{ (\sigma^\beta (\ga_{\kv'\omega})^2 \sigma^\gamma
  -\sigma^\gamma (\ga_{\kv'\omega})^2 \sigma^\beta) 
   \ga_{\kv\omega}\sigma^\delta \ga_{\kv\omega} 
  -{\rm c.c.} \rt\}
\rt. \nonumber\\
&  -\frac{1}{\Omz}
\lt( f\lt(\omega-\frac{\Omz}{2}\rt)
   (\sigma^\beta \ga_{\kv'\omega}\sigma^\gamma
    +\sigma^\gamma \ga_{\kv'\omega}\sigma^\beta)
   \ga_{\kv\omega} \sigma^\delta \ga_{\kv\omega} 
    \rt. \nonumber\\
&  \lt.\lt.\lt.
   -f\lt(\omega+\frac{\Omz}{2}\rt)
   (\sigma^\beta \gr_{\kv'\omega}\sigma^\gamma
    +\sigma^\gamma \gr_{\kv'\omega}\sigma^\beta)
   \gr_{\kv\omega} \sigma^\delta \gr_{\kv\omega} \rt)
\rt]\rt]
, \label{Ddiag2}
\end{align}
where 
$f(\omega)\equiv(e^{\beta\omega}+1)^{-1}$.
Similarly, the third contribution in Fig. \ref{FIGdiag_sf} is obtained as
\begin{align}
{D_{\mu\nu}^{\beta\gamma\delta} }^{(3)}
& =
\lim_{\Omz\ra0}  \sumom \sum_{\kv\kv'} 
\delta_{\mu\nu} \tr\lt[
f'(\omega)
\lt\{ \sigma^\beta \gr_{\kv'}\sigma^\gamma    +\sigma^\gamma \ga_{\kv'}\sigma^\beta \rt\}
\gr_{\kv}\sigma^\delta \ga_{\kv}
    \rt.\nonumber\\
&
 +\lt[ 
-\frac{f(\omega)}{2} 
\lt\{ (\sigma^\beta (\ga_{\kv'\omega})^2 \sigma^\gamma
  -\sigma^\gamma (\ga_{\kv'\omega})^2 \sigma^\beta) 
   \ga_{\kv\omega}\sigma^\delta \ga_{\kv\omega} 
  -{\rm c.c.}  
\rt. \rt. \nonumber\\
& \lt.
(\sigma^\beta \ga_{\kv'\omega} \sigma^\gamma
  +\sigma^\gamma \ga_{\kv'\omega} \sigma^\beta) 
   (\ga_{\kv\omega}\sigma^\delta (\ga_{\kv\omega})^2  
       -(\ga_{\kv\omega})^2\sigma^\delta \ga_{\kv\omega} )  -{\rm c.c.}  
\rt\} \nonumber\\
&  +\frac{1}{\Omz}
\lt( f\lt(\omega-\frac{\Omz}{2}\rt)
   (\sigma^\beta \ga_{\kv'\omega}\sigma^\gamma
    +\sigma^\gamma \ga_{\kv'\omega}\sigma^\beta)
   \ga_{\kv\omega} \sigma^\delta \ga_{\kv\omega} 
    \rt. \nonumber\\
&  \lt.\lt.\lt.
  -f\lt(\omega+\frac{\Omz}{2}\rt)
   (\sigma^\beta \gr_{\kv'\omega}\sigma^\gamma
    +\sigma^\gamma \gr_{\kv'\omega}\sigma^\beta)
   \gr_{\kv\omega} \sigma^\delta \gr_{\kv\omega} \rt)
\rt]\rt]
. \label{Ddiag3}
\end{align}
Noting that only antisymmetric part with respect to $\beta$ and $\gamma$ contribute to the torque, these contributions are summed to be 
\begin{align}
D_{\mu\nu}^{\beta\gamma\delta} & = 
-i \sumom \sum_{\kv\kv'} 
f'(\omega) 
\tr\lt[ (\sigma^\beta \Im(\ga_{\kv'}) \sigma^\gamma
         -\sigma^\gamma \Im(\ga_{\kv'}) \sigma^\beta)
\rt.\nonumber\\
 &  \lt. \times   \lt(
 \frac{k_\mu k_\nu}{m}
(|\gr_{\kv}|^2 \sigma^\delta \ga_{\kv}
  +\gr_{\kv} \sigma^\delta |\gr_{\kv}|^2)   
  + \delta_{\mu\nu}   
 (\gr_{\kv} \sigma^\delta \ga_{\kv}) \rt)
\rt] ,
\end{align}
where $\gr_{\kv}\equiv \gr_{\kv,\omega=0}$ etc.
We see that spin flip processes contributes as additional lifetime as indicated by the imaginary part of spin scattered electron Green's function, $\Im \ga_{\kv'}$.

To estimete the trace in the spin space, we use general identities which hold for $2\times2$ diagonal matrices $B, C$, and $D$ 
(containing only $\sigma_z$ and the identity matrix):
\begin{eqnarray}
 \tr[
(\sigma^\beta B \sigma^\gamma-\sigma^\gamma B \sigma^\beta)
 (C\sigma^\delta D+D\sigma^\delta C)] 
&=&
 2i 
[(\epsilon_{\beta\gamma\delta}-\epsilon_{\beta\gamma z}\delta_{\delta z}) 
\nonumber\\
&& \times 
((BC)_{+}D_{-}+(BC)_{-}D_{+}
+(BD)_{+}C_{-}+(BD)_{-}C_{+}) 
\nonumber\\
& &
+2\epsilon_{\beta\gamma z}\delta_{\delta z}
(B_{+}(CD)_{-}+B_{-}(CD)_{+})] 
\nonumber\\
 \tr[
(\sigma^\beta B \sigma^\gamma-\sigma^\gamma B \sigma^\beta)
 (C\sigma^\delta D-D\sigma^\delta C)] 
&= &
 2(\delta_{\gamma z}\delta_{\beta\delta}
  -\delta_{\beta z}\delta_{\gamma\delta})
\nonumber\\
&& \times ((BC)_{+}D_{-}-(BC)_{-}D_{+}
-(BD)_{+}C_{-}+(BD)_{-}C_{+}) , \nonumber\\
\end{eqnarray}
where the components $B_{\pm}$ are defined as
$B=( B_++B_- +(B_+-B_-) \sigma_z )/2$, etc. 
The result for $D_{\mu\nu}^{\beta\gamma\delta}$ is then obtained as
\begin{align}
D_{\mu\nu}^{\beta\gamma\delta} & =
\delta_{\mu\nu} \lt(
a (\epsilon_{\beta\gamma\delta}-\epsilon_{\beta\gamma z}\delta_{\delta z})
+b (\delta_{\beta\delta}\delta_{\gamma z}
   -\delta_{\gamma\delta}\delta_{\beta z}) \rt),
\end{align}
where the coefficients are given by
\begin{align}
a & =   
-\frac{1}{2\pi}
\sum_{\kv\kv'}\sum_{\sigma\sigma'}
\lt[ \frac{k^2}{3m}
|\gr_{\kv\sigma}|^2 (\ga_{\kv,-\sigma}+\gr_{\kv,-\sigma})
+ ( \gr_{\kv\sigma} \ga_{\kv,-\sigma}+\ga_{\kv\sigma} \gr_{\kv,-\sigma} )  \rt]
(\Im \ga_{\kv'\sigma'}) 
  \nonumber\\
b & =
-\frac{1}{2\pi} \sum_{\kv\kv'}\sum_{\sigma\sigma'}
(i \sigma) 
\ga_{\kv \sigma}\gr_{\kv,-\sigma}
 (\Im \ga_{\kv'\sigma'}).
\end{align}
Using 
$F_{\alpha\beta\gamma}=
-\epsilon_{\alpha\beta\gamma}
-2\sum_{\delta} m_\delta
(\epsilon_{\alpha\gamma\delta}m_\beta 
 -\epsilon_{\alpha\beta\delta}m_\gamma)$, and
$\Av_\mu=\hf (\nv\times\partial_\mu\nv) -A_\mu^z \nv$\cite{TKS08}, 
the torque due to spin-flip  
is obtained as
\begin{align}
{\cal T}_{\rm sf} 
&=  -\frac{2e}{3m}\vs^2 \overline{\Simp}^2 \sum_\mu E_{\mu} 
(a (\nv\times\partial_\mu\nv) -  b \partial_{\mu}\nv).
\end{align}
The coefficients  $a$ and $b$ are calculated as
$a= \pi ({m}/{e^2 \spol})(\sigma_+-\sigma_-)(\DOS_++\DOS_-)$
 and
$b=O(a\times(\ef\tau)^{-1})\simeq 0$, where $\DOS_\pm$ and 
$\sigma_{\sigma}=e^2n_\sigma\tau_\sigma/m$ are the  spin-resolved conductivity and density of states. 
(Coefficient $b$ is treated as zero within the present approximation.)
Therefore, the torque induced by the spin relaxation is simply a $\beta$ term given by   
\begin{align}
{\cal T}_{\rm sf} 
&= -\betasf \frac{P}{e} (\nv\times(\jv\cdot\nabla)\nv),
\label{torquebeta}
\end{align}
where $P\equiv(\sigma_+-\sigma_-)/(\sigma_++\sigma_-)$ is the spin polarization of the current and 
\begin{align}
\betasf 
&=  \frac{2\pi}{3\spol}\nimp \vs^2 \overline{\Simp}^2 (\DOS_++\DOS_-) . 
\label{betasfres}
\end{align}
Defining the spin-flip lifetime ($\tau_s$ of Ref. \cite{KTS06})
as (note that $\overline{S_z}^2+\overline{S_\perp}^2$ of Ref. \cite{KTS06} corresponds to $\frac{2}{3}\overline{\Simp}^2$ here)
${\tausf} ^{-1}=({4\pi}/{3})\nimp \vs^2 \overline{\Simp}^2 (\DOS_++\DOS_-)$, we find 
$\betasf={\hbar}/({2\spol \tausf})$, which agrees with results obtained in Refs. \cite{KTS06,Kohno07}.

\subsection{Torque from spin-orbit interaction}

The torque from spin-orbit interaction, Eq. (\ref{Tso}), is calculated in a similar way.
The spin-orbit interaction is written in the rotated frame as
\begin{align}
\Hso &=  \lamso \intx  \sum_{ijkl}
\epsilon_{ijk} \nabla_j \Vso^{(i)}(\xv)R_{il}(\xv)
\lt(
-\frac{i}{2} \adag \sigma^l \nablarl_k a +A_k^l \adag a
\rt).
\end{align}
The spin-orbit contributions to the spin current and the electron density in the rotated frame are obtained as
\begin{align}
\tilde{\js}_\nu^\rho(\xv) &= 
-\frac{i}{2m} \lamso \sum_{ijkl} \epsilon_{ijk} (\nabla^{\xv}-\nabla^{\xv'})_\nu \int d^3x_1 \nabla_j \Vso^{(i)}(\xv_1) R_{il}(\xv_1) \nonumber\\
& \times 
\tr \lt[\sigma^\rho \Gtil^{(0)}_{\xv,\xv_1'}
\lt(-\frac{i}{2}(\nablar^{\xv_1}-\nablal^{\xv_1'})_k \sigma^l
 +A_k^l(\xv_1)\rt)  
\Gtil^{(0)}_{\xv_1,\xv'}\rt]^<_{\xv'\ra\xv,\xv_1'\ra\xv_1}
\nonumber\\
\nel(\xv) &= 
-\frac{i}{2}\lamso  \sum_{ijkl} \epsilon_{ijk}   ,\int d^3x_1 \nabla_j \Vso^{(i)}(\xv_1) R_{il}(\xv_1)(\nablar^{\xv_1}-\nablal^{\xv_1'})_k 
 \tr \lt[ \Gtil^{(0)}_{\xv,\xv_1'}
\sigma^l \Gtil^{(0)}_{\xv_1,\xv'}\rt]^<_{\xv'\ra\xv,\xv_1'\ra\xv_1}
+O(A).
\end{align}
The torque is then calculated as
\begin{align}
{\cal T}^\alpha_{\rm so}(\xv) &= 
-i \lamso^2   \sum_{\beta\mu\nu\tau} \sum_{jklm} \epsilon_{\alpha\mu\tau} \epsilon_{lm\tau}
  \epsilon_{\nu jk}
\int d^3x_1 R_{\mu\beta}(\xv) R_{\nu \gamma}(\xv_1)  
\sum_{\kv\kv'\pv}\sum_{\kv_1 \kv_1'}
p_l p_j e^{-i\pv\cdot(\xv-\xv_1)}
e^{-i(\kv-\kv')\cdot \xv}
e^{-i(\kv_1-\kv_1')\cdot \xv_1} 
\nonumber\\
& \times
 \average{
 {\Vso}^{(\beta)} (\pv) {\Vso}^{(\tau)} ({-\pv})} \lt[ \frac{1}{2} (k+k')_m (k_1+k_1')_k 
\tr \lt[\sigma^\beta \Gtil^{(0)}_{\kv,\kv_1}
 \sigma^\gamma \Gtil^{(0)}_{\kv_1',\kv'}\rt]^<  \rt. \nonumber\\  
&   \lt.
+ (k+k')_m A_k^\gamma (\xv_1) 
\tr \lt[\sigma^\beta \Gtil^{(0)}_{\kv,\kv_1} \Gtil^{(0)}_{\kv_1',\kv'}\rt]^<   
+ (k_1+k_1')_k A_m^\beta(\xv_1) 
\tr \lt[\Gtil^{(0)}_{\kv,\kv_1} \sigma^\gamma \Gtil^{(0)}_{\kv_1',\kv'}\rt]^<   
\rt].
\end{align}
In the adiabatic limit we consider, 
Green's functions are diagonal in wave vectors,
$\Gtil^{(0)}_{\kv,\kv'}=\delta_{\kv,\kv'}\Gtil^{(0)}_{\kv}$, and the integration over $\xv_1$ can be carried out treating the slowly varying variables
$R(\xv_1)$ and $A(\xv_1)$ as constants, resulting in 
$\int d\xv_1 e^{-i(\pv-\kv+\kv')\cdot(\xv-\xv_1)}
=V\delta_{\pv,\kv-\kv'}$.
We therefore obtain 
\begin{align}
{\cal T}^\alpha_{\rm so}(\xv) &= 
-i \lamso^2  \sum_{\beta\mu\nu\tau} \sum_{jklm}  \epsilon_{\alpha\mu\tau} \epsilon_{lm\tau}
  \epsilon_{\nu jk}
 R_{\mu\beta}(\xv) R_{\nu \gamma}(\xv)  
\sum_{\kv\kv'}
(k-k')_l (k-k')_j 
\nonumber\\
& \times
\average{
{\Vso}^{(\nu)} ({\kv-\kv'}) {\Vso}^{(\tau)} ({-\kv+\kv'})}
\lt[ \frac{1}{2} (k+k')_m (k_1+k_1')_k 
\tr \lt[\sigma^\beta \Gtil^{(0)}_{\kv}
 \sigma^\gamma \Gtil^{(0)}_{\kv'}\rt]^<  \rt. \nonumber\\  
&   \lt.
+ (k+k')_m A_k^\gamma (\xv) 
\tr \lt[\sigma^\beta \Gtil^{(0)}_{\kv} \Gtil^{(0)}_{\kv'}\rt]^< 
+ (k+k')_k A_m^\beta(\xv) 
\tr \lt[\Gtil^{(0)}_{\kv} \sigma^\gamma \Gtil^{(0)}_{\kv'}\rt]^<   
\rt]. \label{tauso2}
\end{align}

We average over spin-orbit impurities so that average remains finite only when the spin polarizations are parallel. 
Impurity averaging is thus given as  
\begin{equation}
\average{{\Vso}^{(\nu)}(\pv) {\Vso} ^{(\tau)}({-\pv'})}_{\rm i}
=\nso \delta_{\nu\tau} \delta_{\pv,\pv'}.
\end{equation}
The result of the torque is
\begin{align}
{\cal T}^\alpha_{\rm so}(\xv) 
&= 
-i \frac{1}{2} \nso\lamso^2
\sum_{\beta\mu\nu\tau} 
\epsilon_{\alpha\mu\nu}  R_{\mu\beta}(\xv) R_{\nu \gamma}(\xv)  
\nonumber\\
& \times
\sum_{\kv\kv'} 
\lt[(\kv\times\kv')_\tau  (\kv\times\kv')_\nu  
\tr \lt[\sigma^\beta \Gtil^{(0)}_{\kv}
 \sigma^\gamma \Gtil^{(0)}_{\kv'}\rt]^<  \rt. \nonumber\\  
&   \lt.
+ (\kv\times\kv')_\tau  [(\kv-\kv')\times \Av^\gamma]_\nu 
\tr \lt[\sigma^\beta \Gtil^{(0)}_{\kv} \Gtil^{(0)}_{\kv'}\rt]^< 
+ (\kv\times\kv')_\nu  [(\kv-\kv')\times \Av^\beta]_\tau 
\tr \lt[\Gtil^{(0)}_{\kv} \sigma^\gamma \Gtil^{(0)}_{\kv'}\rt]^<   
\rt]_{\nu=\tau}.  \label{tauso3} 
\end{align}
The last two terms lead to vanishing contribution in the adiabatic limit.
In fact, these are already linear in $A$ and so $\Gtil^{(0)}$
does not contain spin-flip conponents, and thus 
$\sigma^z$ and $\Gtil^{(0)}$ commute
each other. We therefore obtain 
\begin{align}
[(\kv-\kv')\times \Av^\gamma]_\nu 
\tr \lt[\sigma^\beta \Gtil^{(0)}_{\kv} \Gtil^{(0)}_{\kv'}\rt]^< + [(\kv-\kv')\times \Av^\beta]_\nu 
\tr \lt[\Gtil^{(0)}_{\kv} \sigma^\gamma \Gtil^{(0)}_{\kv'}\rt]^<   \nonumber \\
=
(\delta_{\beta,z}[(\kv-\kv')\times \Av^\gamma]_\nu 
+ \delta_{\gamma,z} [(\kv-\kv')\times \Av^\beta]_\nu )
\tr \lt[\sigma^z \Gtil^{(0)}_{\kv} \Gtil^{(0)}_{\kv'} \rt]^< .
\end{align}  
This contribution is symmetric with respect to $\beta$ and $\gamma$, and results in zero when multiplied by $F_{\mu\nu}^{\alpha\beta\gamma}$, which is asymmetric with respect to $\beta$ and $\gamma$.

The first term of \Eqref{tauso3} can be simplified by using the rotational symmetry of electron,
$\average{ (\kv\times\kv')_\tau  (\kv\times\kv')_\tau}
 =\frac{1}{3} \average{(\kv\times\kv')\cdot(\kv\times\kv')}
 =\frac{1}{3} \average{(k^2 {k'} ^2-(\kv\cdot\kv')^2)}$ ($\average{\ }$ denotes the angular average), as
\begin{align}
{\cal T}^\alpha_{\rm so}(\xv) &= 
-i \frac{1}{6} \nso\lamso^2 \sum_{\beta\mu\nu\tau} \sum_{\kv\kv'}
F_{\mu\nu}^{\alpha\beta\gamma}  
 (k^2 {k'} ^2-(\kv\cdot\kv')^2)
\tr \lt[\sigma^\beta \Gtil^{(0)}_{\kv}
 \sigma^\gamma \Gtil^{(0)}_{\kv'}\rt]^< .  \label{tauso4} 
\end{align}

We therefore see that the expression is similar to that of spin-flip impurity case, \Eqref{Tsf1}.
Including the effect of electric field and gauge field to linear oder in both similarly to the spin flip impurity case, we obtain the torque as 
\begin{align}
{\cal T}_{\rm so} 
&=  -\frac{e}{6m}\nso {\lamso}^2 a'  \sum_\mu E_{\mu} 
(\nv\times\partial_\mu\nv) ,
\end{align}
where coefficient is given as
\begin{align}
a' & =   
-\frac{1}{2\pi}
\sum_{\kv\kv'}\sum_{\sigma\sigma'} 
(k^2 {k'} ^2-(\kv\cdot\kv')^2)
\lt[ \frac{k^2}{3m}
|\gr_{\kv\sigma}|^2 (\ga_{\kv,-\sigma}+\gr_{\kv,-\sigma})
+ ( \gr_{\kv\sigma} \ga_{\kv,-\sigma}+\ga_{\kv\sigma} \gr_{\kv,-\sigma} )  \rt]
(\Im \ga_{\kv'\sigma'}) .
\end{align}

The coefficient is calculated as
$a'= \pi \frac{2m}{3e^2 \spol}(\sigma_+ \kfu^2 -\sigma_-\kfd ^2)(\DOS_+\kfu^2+\DOS_-\kfd^2)$.
Therefore, spin-orbit interaction yields the $\beta$ term with coefficient given by 
\begin{align}
\betaso 
&=  \frac{1}{2\spol}\frac{1}{n_+\tau_+ - n_-\tau_-} \lt(\frac{n_+\tau_+}{\tauso_+}-\frac{n_-\tau_-}{\tauso_-}\rt),
\label{betasores}
\end{align}
where
\begin{align}
\frac{1}{\tauso_{\pm}} 
&\equiv  \frac{2\pi}{9} \nso\lamso^2 k_{F\pm}^2 
  (\DOS_+\kfu^2+\DOS_-\kfd^2),
\end{align}
with ${\tauso_{\pm}}$ as the lifetime due to spin-orbit interaction.

The total current-induced torque in the adiabatic limit is 
therefore given by 
Eqs. (\ref{stt}) (\ref{torquebeta}) (\ref{betasfres}) (\ref{betasores}) as
\begin{align}
\torquev
&= -\frac{P}{2e}(\nabla\cdot\jv) \nv
- \betasr \frac{P}{e} (\nv\times(\jv\cdot\nabla)\nv),
\end{align}
with $\betasr\equiv \betasf+\betaso$.

\section{Summary}

In summary, we demonstrated that the spin continuity equation represents the current-induced torques acting on the magnetization, and that it can be used for microscopic determination of the torques. 
The present formalism does not assume separation of magnetization and conduction electron degrees of freedom and can directly be applied to itinerant electron systems without mean-field approximation.
In this paper, the formalism was applied to the $s$-$d$ model in the presence of spin relaxation caused due to spin-flip scattering and spin-orbit interaction with impurities.
Both relaxation processes were shown to induce the so called $\beta$ torque term.

Application of the formalism to realistic itinerant system using first principles calculations would be very interesting, since it would allow for quantitative estimations of current-induced switching.
Of particular interest are the systems with enhanced spin-orbit interaction near surfaces and multilayers. 
Our formulation can be easliy extended to describe these systems.

Further improvement of the present theory would be to include effects caused by electron-electron correlation.
If the correlation is represented within the mean-field approximation by a local spin-dependent potential, the torque is straightforwardly calculated similarly to the estimate of spin-flip scattering. 
Treatment beyond mean-field would be an important furture work.

\begin{acknowledgments}
The authors thank H. Akai, M. Ogura and H. Kohno for valuable discussions.
G. T. acknowledges Grant-in-Aid for Scientific Reseach on Priority Areas for financial support.
P. E. thanks the SFB491 and the DFG for financial support. 
\end{acknowledgments}

\bibliography{/home/tatara/References/dw08}

\begin{thebibliography}{25}
\expandafter\ifx\csname natexlab\endcsname\relax\def\natexlab#1{#1}\fi
\expandafter\ifx\csname bibnamefont\endcsname\relax
  \def\bibnamefont#1{#1}\fi
\expandafter\ifx\csname bibfnamefont\endcsname\relax
  \def\bibfnamefont#1{#1}\fi
\expandafter\ifx\csname citenamefont\endcsname\relax
  \def\citenamefont#1{#1}\fi
\expandafter\ifx\csname url\endcsname\relax
  \def\url#1{\texttt{#1}}\fi
\expandafter\ifx\csname urlprefix\endcsname\relax\def\urlprefix{URL }\fi
\providecommand{\bibinfo}[2]{#2}
\providecommand{\eprint}[2][]{\url{#2}}

\bibitem[{\citenamefont{Tatara and Kohno}(2004)}]{TK04}
\bibinfo{author}{\bibfnamefont{G.}~\bibnamefont{Tatara}} \bibnamefont{and}
  \bibinfo{author}{\bibfnamefont{H.}~\bibnamefont{Kohno}},
  \bibinfo{journal}{Phys. Rev. Lett.} \textbf{\bibinfo{volume}{92}},
  \bibinfo{eid}{086601} (\bibinfo{year}{2004}).

\bibitem[{\citenamefont{Zhang and Li}(2004)}]{Zhang04}
\bibinfo{author}{\bibfnamefont{S.}~\bibnamefont{Zhang}} \bibnamefont{and}
  \bibinfo{author}{\bibfnamefont{Z.}~\bibnamefont{Li}}, \bibinfo{journal}{Phys.
  Rev. Lett.} \textbf{\bibinfo{volume}{93}}, \bibinfo{eid}{127204}
 (\bibinfo{year}{2004}).

\bibitem[{\citenamefont{Waintal and Viret}(2004)}]{Waintal04}
\bibinfo{author}{\bibfnamefont{X.}~\bibnamefont{Waintal}} \bibnamefont{and}
  \bibinfo{author}{\bibfnamefont{M.}~\bibnamefont{Viret}},
  \bibinfo{journal}{Europhys. Lett.} \textbf{\bibinfo{volume}{65}},
  \bibinfo{pages}{427} (\bibinfo{year}{2004}).

\bibitem[{\citenamefont{Kohno et~al.}(2006)\citenamefont{Kohno, Tatara, and
  Shibata}}]{KTS06}
\bibinfo{author}{\bibfnamefont{H.}~\bibnamefont{Kohno}},
  \bibinfo{author}{\bibfnamefont{G.}~\bibnamefont{Tatara}}, \bibnamefont{and}
  \bibinfo{author}{\bibfnamefont{J.}~\bibnamefont{Shibata}},
  \bibinfo{journal}{J. Phys. Soc. Jpn.} \textbf{\bibinfo{volume}{75}},
  \bibinfo{pages}{113706} (\bibinfo{year}{2006}).

\bibitem[{\citenamefont{Kohno and Shibata}(2007)}]{Kohno07}
\bibinfo{author}{\bibfnamefont{H.}~\bibnamefont{Kohno}} \bibnamefont{and}
  \bibinfo{author}{\bibfnamefont{J.}~\bibnamefont{Shibata}},
  \bibinfo{journal}{J. Phys. Soc. Jpn.} \textbf{\bibinfo{volume}{76}},
  \bibinfo{pages}{063710} (\bibinfo{year}{2007}).

\bibitem[{\citenamefont{Tatara et~al.}(2008{\natexlab{a}})\citenamefont{Tatara,
  Kohno, and Shibata}}]{Tatara_cd08}
\bibinfo{author}{\bibfnamefont{G.}~\bibnamefont{Tatara}},
  \bibinfo{author}{\bibfnamefont{H.}~\bibnamefont{Kohno}}, \bibnamefont{and}
  \bibinfo{author}{\bibfnamefont{J.}~\bibnamefont{Shibata}},
  \bibinfo{journal}{cond-mat arXiv:0807.2894v1}
  (\bibinfo{year}{2008}{\natexlab{a}}).

\bibitem[{\citenamefont{Berger}(1978)}]{Berger78}
\bibinfo{author}{\bibfnamefont{L.}~\bibnamefont{Berger}}, \bibinfo{journal}{J.
  Appl. Phys.} \textbf{\bibinfo{volume}{49}}, \bibinfo{pages}{2156}
  (\bibinfo{year}{1978}).

\bibitem[{\citenamefont{Berger}(1984)}]{Berger84}
\bibinfo{author}{\bibfnamefont{L.}~\bibnamefont{Berger}}, \bibinfo{journal}{J.
  Appl. Phys.} \textbf{\bibinfo{volume}{55}}, \bibinfo{pages}{1954}
  (\bibinfo{year}{1984}).

\bibitem[{\citenamefont{Slonczewski}(1996)}]{Slonczewski96}
\bibinfo{author}{\bibfnamefont{J.~C.} \bibnamefont{Slonczewski}},
  \bibinfo{journal}{J. Magn Magn Mater.} \textbf{\bibinfo{volume}{159}},
  \bibinfo{pages}{L1} (\bibinfo{year}{1996}).

\bibitem[{\citenamefont{Tserkovnyak et~al.}(2006)\citenamefont{Tserkovnyak,
  Skadsem, Brataas, and Bauer}}]{Tserkovnyak06}
\bibinfo{author}{\bibfnamefont{Y.}~\bibnamefont{Tserkovnyak}},
  \bibinfo{author}{\bibfnamefont{H.~J.} \bibnamefont{Skadsem}},
  \bibinfo{author}{\bibfnamefont{A.}~\bibnamefont{Brataas}}, \bibnamefont{and}
  \bibinfo{author}{\bibfnamefont{G.~E.~W.} \bibnamefont{Bauer}},
  \bibinfo{journal}{Phys. Rev. B} \textbf{\bibinfo{volume}{74}},
  \bibinfo{eid}{144405} (\bibinfo{year}{2006}).

\bibitem[{\citenamefont{Tatara et~al.}(2007)\citenamefont{Tatara, Kohno,
  Shibata, Lemaho, and Lee}}]{TKSLL07}
\bibinfo{author}{\bibfnamefont{G.}~\bibnamefont{Tatara}},
  \bibinfo{author}{\bibfnamefont{H.}~\bibnamefont{Kohno}},
  \bibinfo{author}{\bibfnamefont{J.}~\bibnamefont{Shibata}},
  \bibinfo{author}{\bibfnamefont{Y.}~\bibnamefont{Lemaho}}, \bibnamefont{and}
  \bibinfo{author}{\bibfnamefont{K.-J.} \bibnamefont{Lee}},
  \bibinfo{journal}{J. Phys. Soc. Jpn.} \textbf{\bibinfo{volume}{76}},
  \bibinfo{pages}{054707} (\bibinfo{year}{2007}).

\bibitem[{\citenamefont{Duine et~al.}(2007)\citenamefont{Duine, nez, Sinova,
  and MacDonald}}]{Duine07}
\bibinfo{author}{\bibfnamefont{R.~A.} \bibnamefont{Duine}},
  \bibinfo{author}{\bibfnamefont{A.~S.} \bibnamefont{Nunez}},
  \bibinfo{author}{\bibfnamefont{J.}~\bibnamefont{Sinova}}, \bibnamefont{and}
  \bibinfo{author}{\bibfnamefont{A.~H.} \bibnamefont{MacDonald}},
  \bibinfo{journal}{Phys. Rev. B} \textbf{\bibinfo{volume}{75}},
  \bibinfo{eid}{214420}  (\bibinfo{year}{2007}).

\bibitem[{\citenamefont{Thorwart and Egger}(2007)}]{Thorwart07}
\bibinfo{author}{\bibfnamefont{M.}~\bibnamefont{Thorwart}} \bibnamefont{and}
  \bibinfo{author}{\bibfnamefont{R.}~\bibnamefont{Egger}},
  \bibinfo{journal}{Phys. Rev. B} \textbf{\bibinfo{volume}{76}},
  \bibinfo{eid}{214418}  (\bibinfo{year}{2007}).

\bibitem[{\citenamefont{Piechon and Thiaville}(2007)}]{Piechon07}
\bibinfo{author}{\bibfnamefont{F.}~\bibnamefont{Piechon}} \bibnamefont{and}
  \bibinfo{author}{\bibfnamefont{A.}~\bibnamefont{Thiaville}},
  \bibinfo{journal}{Phys. Rev. B} \textbf{\bibinfo{volume}{75}},
  \bibinfo{eid}{174414}  (\bibinfo{year}{2007}).

\bibitem[{\citenamefont{Tatara et~al.}(2008{\natexlab{b}})\citenamefont{Tatara,
  Kohno, and Shibata}}]{TKS08}
\bibinfo{author}{\bibfnamefont{G.}~\bibnamefont{Tatara}},
  \bibinfo{author}{\bibfnamefont{H.}~\bibnamefont{Kohno}}, \bibnamefont{and}
  \bibinfo{author}{\bibfnamefont{J.}~\bibnamefont{Shibata}},
  \bibinfo{journal}{J. Phys. Soc. Jpn.} \textbf{\bibinfo{volume}{77}},
  \bibinfo{pages}{031003} (\bibinfo{year}{2008}{\natexlab{b}}).

\bibitem[{\citenamefont{Thiaville et~al.}(2005)\citenamefont{Thiaville,
  Nakatani, Miltat, and Suzuki}}]{Thiaville05}
\bibinfo{author}{\bibfnamefont{A.}~\bibnamefont{Thiaville}},
  \bibinfo{author}{\bibfnamefont{Y.}~\bibnamefont{Nakatani}},
  \bibinfo{author}{\bibfnamefont{J.}~\bibnamefont{Miltat}}, \bibnamefont{and}
  \bibinfo{author}{\bibfnamefont{Y.}~\bibnamefont{Suzuki}},
  \bibinfo{journal}{Europhys. Lett.} \textbf{\bibinfo{volume}{69}},
  \bibinfo{pages}{990} (\bibinfo{year}{2005}).

\bibitem[{\citenamefont{Tatara et~al.}(2006)\citenamefont{Tatara, Takayama,
  Kohno, Shibata, Nakatani, and Fukuyama}}]{TTKSNF06}
\bibinfo{author}{\bibfnamefont{G.}~\bibnamefont{Tatara}},
  \bibinfo{author}{\bibfnamefont{T.}~\bibnamefont{Takayama}},
  \bibinfo{author}{\bibfnamefont{H.}~\bibnamefont{Kohno}},
  \bibinfo{author}{\bibfnamefont{J.}~\bibnamefont{Shibata}},
  \bibinfo{author}{\bibfnamefont{Y.}~\bibnamefont{Nakatani}}, \bibnamefont{and}
  \bibinfo{author}{\bibfnamefont{H.}~\bibnamefont{Fukuyama}},
  \bibinfo{journal}{J. Phys. Soc. Jpn.} \textbf{\bibinfo{volume}{75}},
  \bibinfo{pages}{64708} (\bibinfo{year}{2006}).

\bibitem[{\citenamefont{Heyne et~al.}(2008)\citenamefont{Heyne, Kl\"aui,
  Backes, Moore, Krzyk, Rudiger, Heyderman, Rodriguez, Nolting, Mentes
  et~al.}}]{Heyne08}
\bibinfo{author}{\bibfnamefont{L.}~\bibnamefont{Heyne}},
  \bibinfo{author}{\bibfnamefont{M.}~\bibnamefont{Kl\"aui}},
  \bibinfo{author}{\bibfnamefont{D.}~\bibnamefont{Backes}},
  \bibinfo{author}{\bibfnamefont{T.~A.} \bibnamefont{Moore}},
  \bibinfo{author}{\bibfnamefont{S.}~\bibnamefont{Krzyk}},
  \bibinfo{author}{\bibfnamefont{U.}~\bibnamefont{Rudiger}},
  \bibinfo{author}{\bibfnamefont{L.~J.} \bibnamefont{Heyderman}},
  \bibinfo{author}{\bibfnamefont{A.~F.} \bibnamefont{Rodriguez}},
  \bibinfo{author}{\bibfnamefont{F.}~\bibnamefont{Nolting}},
  \bibinfo{author}{\bibfnamefont{T.~O.} \bibnamefont{Mentes}},
  \bibnamefont{et~al.}, \bibinfo{journal}{Phys. Rev. Lett.}
  \textbf{\bibinfo{volume}{100}}, \bibinfo{eid}{066603}
   (\bibinfo{year}{2008}).

\bibitem[{\citenamefont{Thomas et~al.}(2006)\citenamefont{Thomas, Hayashi,
  Jiang, Moriya, Rettner, and Parkin}}]{Thomas06}
\bibinfo{author}{\bibfnamefont{L.}~\bibnamefont{Thomas}},
  \bibinfo{author}{\bibfnamefont{M.}~\bibnamefont{Hayashi}},
  \bibinfo{author}{\bibfnamefont{X.}~\bibnamefont{Jiang}},
  \bibinfo{author}{\bibfnamefont{R.}~\bibnamefont{Moriya}},
  \bibinfo{author}{\bibfnamefont{C.}~\bibnamefont{Rettner}}, \bibnamefont{and}
  \bibinfo{author}{\bibfnamefont{S.~S.~P.} \bibnamefont{Parkin}},
  \bibinfo{journal}{Nature} \textbf{\bibinfo{volume}{443}},
  \bibinfo{pages}{197} (\bibinfo{year}{2006}).

\bibitem[{\citenamefont{Murakami et~al.}(2004)\citenamefont{Murakami, Nagosa,
  and Zhang}}]{Murakami04}
\bibinfo{author}{\bibfnamefont{S.}~\bibnamefont{Murakami}},
  \bibinfo{author}{\bibfnamefont{N.}~\bibnamefont{Nagosa}}, \bibnamefont{and}
  \bibinfo{author}{\bibfnamefont{S.-C.} \bibnamefont{Zhang}},
  \bibinfo{journal}{Phys. Rev. B} \textbf{\bibinfo{volume}{69}},
  \bibinfo{pages}{235206} (\bibinfo{year}{2004}).

\bibitem[{\citenamefont{Culcer et~al.}(2004)\citenamefont{Culcer, Sinova,
  Sinitsyn, Jungwirth, MacDonald, and Niu}}]{Culcer04}
\bibinfo{author}{\bibfnamefont{D.}~\bibnamefont{Culcer}},
  \bibinfo{author}{\bibfnamefont{J.}~\bibnamefont{Sinova}},
  \bibinfo{author}{\bibfnamefont{N.~A.} \bibnamefont{Sinitsyn}},
  \bibinfo{author}{\bibfnamefont{T.}~\bibnamefont{Jungwirth}},
  \bibinfo{author}{\bibfnamefont{A.~H.} \bibnamefont{MacDonald}},
  \bibnamefont{and} \bibinfo{author}{\bibfnamefont{Q.}~\bibnamefont{Niu}},
  \bibinfo{journal}{Phys. Rev. Lett.} \textbf{\bibinfo{volume}{93}},
  \bibinfo{pages}{046602} (\bibinfo{year}{2004}).

\bibitem[{\citenamefont{Shi et~al.}(2006)\citenamefont{Shi, Zhang, Xiao, and
  Niu}}]{Shi06}
\bibinfo{author}{\bibfnamefont{J.}~\bibnamefont{Shi}},
  \bibinfo{author}{\bibfnamefont{P.}~\bibnamefont{Zhang}},
  \bibinfo{author}{\bibfnamefont{D.}~\bibnamefont{Xiao}}, \bibnamefont{and}
  \bibinfo{author}{\bibfnamefont{Q.}~\bibnamefont{Niu}},
  \bibinfo{journal}{Phys. Rev. Lett.} \textbf{\bibinfo{volume}{96}},
  \bibinfo{eid}{076604}  (\bibinfo{year}{2006}).

\bibitem[{\citenamefont{Zhang et~al.}(2008)\citenamefont{Zhang, Wang, Shi,
  Xiao, and Niu}}]{Zhang08}
\bibinfo{author}{\bibfnamefont{P.}~\bibnamefont{Zhang}},
  \bibinfo{author}{\bibfnamefont{Z.}~\bibnamefont{Wang}},
  \bibinfo{author}{\bibfnamefont{J.}~\bibnamefont{Shi}},
  \bibinfo{author}{\bibfnamefont{D.}~\bibnamefont{Xiao}}, \bibnamefont{and}
  \bibinfo{author}{\bibfnamefont{Q.}~\bibnamefont{Niu}},
  \bibinfo{journal}{Phys. Rev. B} \textbf{\bibinfo{volume}{77}},
  \bibinfo{eid}{075304}  (\bibinfo{year}{2008}).


\bibitem[{\citenamefont{Caroli et~al.}(1971)\citenamefont{Caroli, Nozieres, and
  Saint-James}}]{Caroli71}
\bibinfo{author}{\bibfnamefont{R.}~\bibnamefont{Caroli},
  \bibfnamefont{C.and~Combescot}},
  \bibinfo{author}{\bibfnamefont{P.}~\bibnamefont{Nozieres}}, \bibnamefont{and}
  \bibinfo{author}{\bibfnamefont{D.}~\bibnamefont{Saint-James}},
  \bibinfo{journal}{J. Phys. C: Solid St. Phys.} \textbf{\bibinfo{volume}{4}},
  \bibinfo{pages}{916} (\bibinfo{year}{1971}).

\bibitem[{\citenamefont{Wang et~al.}(2008)\citenamefont{Wang, Xu, and
  Xia}}]{Wang08}
\bibinfo{author}{\bibfnamefont{S.}~\bibnamefont{Wang}},
  \bibinfo{author}{\bibfnamefont{Y.}~\bibnamefont{Xu}}, \bibnamefont{and}
  \bibinfo{author}{\bibfnamefont{K.}~\bibnamefont{Xia}},
  \bibinfo{journal}{Phys. Rev. B} \textbf{\bibinfo{volume}{77}},
  \bibinfo{eid}{184430} (\bibinfo{year}{2008}).

\end{thebibliography}
\end{document}